\documentclass[aps,pre,showpacs,superscriptaddress,twocolumn,floatfix]{revtex4}
\usepackage{graphicx}
\usepackage{amsmath}
\newcommand{\rt}{\rightarrow}
\newcommand{\be}{\begin{equation}}
\newcommand{\ee}{\end{equation}}
\newcommand{\bea}{\begin{eqnarray}}
\newcommand{\eea}{\end{eqnarray}}

\begin{document}

\title{Coarse grained dynamics of the freely cooling granular gas 
in one dimension}

\author{Mahendra Shinde}

\email{mahendra.statp@gmail.com}

\affiliation{Beijing National Laboratory for Condensed Matter Physics, 
Institute of Physics, Chinese Academy of Sciences, Beijing-100080, China
}
\affiliation{Centre for Nonlinear Studies and the Beijing-Hong
Kong-Singapore Joint Centre for Nonlinear and Complex Systems,
Hong Kong Baptist University, Kowloon Tong, Hong Kong, China}

\author{Dibyendu Das}

\email{dibyendu@phy.iitb.ac.in}

\affiliation{Department of Physics, Indian Institute of Technology, 
Bombay, Powai, Mumbai-400076, India}

\author{R. Rajesh}

\email{rrajesh@imsc.res.in}

\affiliation{Institute of Mathematical Sciences, CIT campus, Taramani, 
Chennai-600113, India}

\date{\today}

\begin{abstract}
We study the dynamics and structure of clusters in the 
inhomogeneous clustered regime of a freely cooling granular gas of 
point particles in one dimension. The coefficient of restitution 
is modeled as $r_0<1$ or $1$ depending on whether the 
relative speed is greater or smaller than a velocity scale $\delta$.
The effective fragmentation rate of a cluster is shown to rise sharply
beyond a $\delta$ dependent time scale. This crossover is coincident with the
velocity fluctuations within a cluster becoming order $\delta$. Beyond this
crossover time, the cluster size distribution develops a nontrivial power law
distribution, whose scaling properties are related to those of the velocity
fluctuations. We argue that these underlying features are responsible 
behind the recently observed nontrivial coarsening behaviour in the 
one dimensional freely
cooling granular gas.
\end{abstract}


\maketitle

\section{Introduction}

Consider a collection of particles, initially distributed randomly in
space, evolving in time through ballistic transport and inelastic
collisions. Such a system has been studied extensively as a simple
model of granular systems as well as a tractable model in
nonequilibrium statistical mechanics
\cite{carnevale,young,goldhirsch1993,frachebourg1999,frachebourg2000,bennaim1,bennaim2,trizac1,
  trizac2,puri,shinde1,shinde2}. At initial times, the system
undergoes homogeneous cooling, the energy decreasing with time $t$ as
$t^{-2}$ in accordance with Haffs law \cite{haff,haffexperi}.  In this
regime, the particles remain homogeneously distributed with
inter-particle spacing being the only relevant length scale.  At times larger
than a crossover time $t_c$, the system crosses over to an
inhomogeneous clustering regime where the energy decreases as
$t^{-\theta}$, where $\theta$ varies with dimension and is different
from $2$ in dimensions lower than the upper critical dimension. In
this regime, there is a growing length scale ${\cal L}_t$, determined
by the size of the largest cluster.

In one dimension, much more is known than in higher dimensions.
Through an exact solution \cite{frachebourg1999,frachebourg2000} of
the problem with coefficient of restitution set to zero (sticky gas),
and extensive simulations \cite{bennaim1} of the inelastic gas, it is
known that $\theta=2/3$. The sticky limit may also be mapped to the
dynamics of shocks in the inviscid
Burgers equation \cite{kida}.  Recently, we showed that, 
when the coefficient of restitution depends on the impact velocity, then
a new timescale $t_1$ further subdivides the inhomogeneous clustering
regime into two sub-regimes \cite{shinde1,shinde2}. This was based on
a study of the density--density and the velocity--velocity correlation
functions.  For times $t_c<t<t_1$, these structure functions scale
exactly as in the sticky gas, obeying what is known as the Porod law
\cite{porod}. However, for times $t>t_1$, the inelastic gas deviates
from the sticky gas limit and the correlation functions violate Porod
law. In addition, the density distribution and inter-particle distance
distribution develop into power laws that are qualitatively different
from that seen in earlier times. We will refer to the two sub-regimes
as the Porod and fluctuation dominated ordering regimes,
respectively.

Although the macroscopic statistical quantities studied in
Refs.~\cite{shinde1,shinde2} establish that the Porod and the fluctuation
dominated ordering regimes are
distinct, they do not reveal how fluctuations start dominating beyond
the time scale $t_1$. It was speculated that the fluctuation
dominated ordering regime
should have an effective process of fragmentation that will compete
with the otherwise strong effective aggregation. Whether coarse
grained density clusters break up or remain 
coherently moving objects can be directly checked by studying their
dynamics. In this paper, we study cluster dynamics, in particular effective
fragmentation rates, and show that the ordering process gets disturbed 
beyond a certain timescale.

The second question is regarding the origin of the crossover timescale
$t_1$.  Earlier this was shown to depend on a velocity scale
$\delta$ associated with the coefficient of restitution
\cite{shinde1,shinde2}.  The coefficient of restitution is often
modelled as a function of the relative velocity-- rather than being a
constant--
such that collisions
become near elastic for relative velocities smaller than
$\delta$. This property is consistent with experiments
\cite{raman,others1,others2}, required by theory
\cite{brilliantovbook}, as well as essential in simulations to prevent
inelastic collapse \cite{young}. In this paper we demonstrate 
that the time $t_1$ is marked by the particle
velocity fluctuations within density clusters becoming of the order
$\delta$. Thus, it is not the typical velocities but rather the velocity 
fluctuations that matter for correlation functions.
We also present a consistent
scaling theory to understand the dependence of velocity fluctuations,
fragmentation rates and cluster size distribution on the parameters
$\delta$, time $t$ and cluster size $m$. 

In Sec.~\ref{sec:model}, we define the model, the quantities
of interest, and give details of the simulation. In Sec.~\ref{sec:results}, we
present results from numerical simulations for velocity fluctuations,
fragmentation rates and cluster size distribution. We develop a scaling theory
which enables us to understanding the scaling of the above quantities in terms
of two exponents.  Sec.~\ref{sec:summary} contains a summary
and discussion of results.

\section{\label{sec:model}Model and definitions}

In this section, we define the microscopic model, and the quantities of
interest.  The model consists of a collection of $N$ point particles 
of equal mass on a
ring of length $L$. Each particle moves ballistically until it collides with
another particle. The collisions conserve momentum, but are inelastic,
such that when two particles
with initial velocities $u_i$ and $u_j$ collide, the final velocities
$u_i^{\prime}$ and $u_j^{\prime}$ are determined by 
\be 
u_{i,j}^{\prime} =
u_{i,j} \left(\frac{1-r}{2} \right) + u_{j,i}
\left(\frac{1+r}{2}\right),  
\ee
where $r$ is the coefficient of restitution.
The coefficient of restitution is velocity dependent such that 
collisions become elastic when the relative velocity tends to zero. As
mentioned in the introduction, this feature is consistent with experiments and
theory, and often used in simulations to circumvent inelastic collapse.
Our results will not depend on the 
detailed dependence of the coefficient of restitution on relative 
velocity: we therefore choose a convenient functional form
\cite{bennaim1,bennaim2},
\be 
r(v_{\rm rel}) = 
\begin{cases}
r_0 & \text{if $v_{\rm rel} > \delta$}, \\
1   & \text{if $v_{\rm rel} \leq \delta$,}
\end{cases}
\label{r}
\ee
where $\delta$ is a velocity scale in the problem. The collisions are
elastic for relative velocities smaller than $\delta$. 
The particles are initially distributed
randomly in space with their velocities drawn from a Gaussian
distribution.

Starting from the above microscopic model, we desire to
study emergent processes such as fragmentation and aggregation of a
collection of particles.  To this end, a coarse grained description
has to be introduced, which we do as follows. 
Divide the ring into $N$ equally sized
boxes. Let the number of particles in the $i^{\rm th}$ ($i=1,2,\ldots,
N$) box be called  the box density. We define a cluster to be a
collection of contiguous boxes with non-zero box density surrounded by
two empty boxes. The total number of particles in this cluster will be
called the size of the cluster. Similar definition has been
used elsewhere, for example see \cite{huepe}. In earlier papers
\cite{shinde1,shinde2}, we had studied velocity-velocity and
density-density correlations using the coarse grained box density.
However, a single particle moving across the boundary of a box
results in change of density. To study clusters, any coarse graining
should make sure that particles move significantly, i.e. by at least a
box spacing ($L/N$) to cause a change in configuration. The above
definition of a cluster has this property.

One of the quantities of interest in this paper is velocity fluctuations
$\sigma$ within a cluster. Let ${\bf u}_c$ be the centre of mass
velocity of a cluster, i.e., ${\bf u}_c=m^{-1} \sum_{i=1}^{m} {\bf u}_i$, 
where ${\bf u}_i$ are the velocities of the particles constituting the
cluster, and $m$ is the size of the cluster.
Then, the velocity fluctuations of that cluster is 
$\sigma(m,t) = m^{-1} \sum_{i=1}^{m} ({\bf u}_{i} - {\bf u}_c)^2$, where the
summation is again over all the particles constituting the cluster. When
$\sigma$ is much smaller than $u_c^2$, then the cluster is compact and
stable with respect to the velocity fluctuations. However,
if they are of same order, or if $\sigma \sim \delta^2$, 
then the cluster may start breaking apart. This
effect can be captured by defining an effective fragmentation rate of a
cluster, as described below.

In the space of cluster sizes, a stochastic dynamics may be defined,
with rates for the different processes being determined from
simulations. As the clusters evolve in time due to the entry and exit of
particles into and out of a cluster, we ask what the rates of
transition from a cluster size $m$ to $m'$ are. 
Let $W(m \rt m';t)$ be the rate  at which a size $m$ changes into a
size $m'$ at time $t$.  If $N(m,t)$ is the number of clusters per unit
lattice site of size $m$ at time $t$, then its time evolution is described by 
an effective Master equation \cite{kampen}:
\bea
\frac{d N(m,t)}{d t} &= & 
\sum_{m'}  W(m'\rt m;t) N(m',t) \nonumber\\
&-& \sum_{m'}  W(m \rt m';t) N(m,t).
\label{master} 
\eea 

If in a time interval $\Delta t$, the cluster size decreases, then the
cluster is said to have undergone fragmentation. 
Thus, an effective fragmentation rate $W_f(m,t)$  
of a cluster of size size $m$ at time $t$ can be defined as
\bea
W_f(m,t) = \sum_{m'<m} W (m \rt m';t).
\label{eq:frag_defn}
\eea
We note that, in the above, fragmentation is  an emergent process, 
not defined apriori in the microscopic dynamics, unlike some other
models of granular gas where fragmentation occurs on collision\cite{Pagona,
hidalgo}.

We study the model by means of event driven molecular dynamics
simulations \cite{rapaportbook}. 
In the simulations the number density $N/L$ is fixed to be 
to one and the number of particles to be $N=20000$.
The results in the paper do not depend on
the precise values of the parameters $r_0$ and $\delta$, as long as $\delta$ is
much smaller than initial velocity differences of adjacent particles. We use
generic values $\delta = 0.001, 0.002, 0.004, 0.008$, and $r_0 = 0.1$
in the simulations. The initial velocities are chosen from a Gaussian
distribution with width $1$. 
The data is typically averaged over $20000$ -- $30000$
different initial conditions. 
All averages will be over space and different histories and will
be denoted by $\langle \cdots \rangle$. Also, we use reduced units in
which all lengths are measured in terms of initial mean inter particle
spacing and times in terms of initial mean collision time.

\section{\label{sec:results}Results}

There are four velocity scales in the problem. First is the typical
speed of a cluster which decreases in time as $t^{-1/3}$
\cite{frachebourg1999,bennaim1}.  Second is
the root mean square velocity fluctuations $\sqrt{\sigma}$ within a
cluster (discussed in detail below). 
Third is $\delta$,
characterising the coefficient of restitution [see
Eq.~(\ref{r})], while the fourth
corresponds to the initial velocity distribution. At large times,
there is no memory of the initial velocity distribution, and it will
play no role in the subsequent discussion. When the typical speeds become of
order $\delta$, then almost all collisions are elastic and energy does not
decrease any more. We will denote the latter crossover time 
by $t_2$. Clearly, $t_2 \sim
\delta^{-3}$. 
It is possible that the velocity fluctuations scale with time
differently from the typical velocity. If so, we have a possibility of a
different crossover time which is marked by the velocity fluctuations becoming
order $\delta$.

\begin{figure}
\includegraphics[width=\columnwidth]{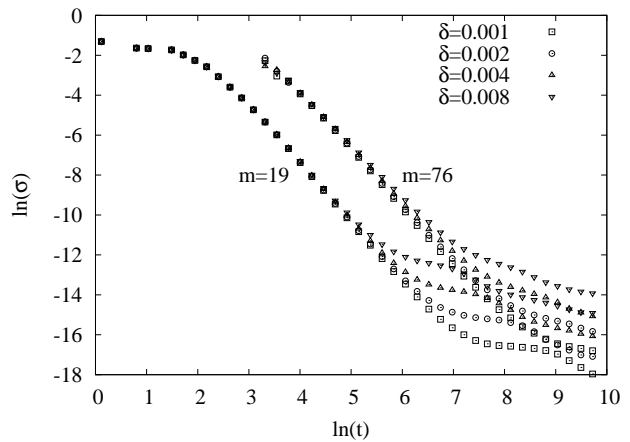}
\caption{\label{sigma_raw} The velocity fluctuations $\sigma$
within a cluster of size $m$ as a function of time $t$ for different
$\delta$. The data are for sizes $m=19$ and $m=76$. For short times the
curves, for a given size, decay as a power law and is independent of
$\delta$. Deviation from power law is seen earlier for larger $\delta$.
}
\end{figure}

We first characterise the velocity fluctuations $\sigma$. In
Fig.~\ref{sigma_raw}, we show the variation of $\sigma$ with time $t$
for different values of $\delta$ and two values of cluster size
$m$. For short times, $\sigma$ is independent of $\delta$, and after
initial transients, decays in time as a power law, with the exponent
independent of $m$ and the prefactor dependent on $m$. At large
times, $\sigma$ deviates from the power law behaviour and is constant
for a while. We argue that this crossover occurs when $\sigma$ is of
order $\delta^2$ --- velocity fluctuations and hence relative
velocities are such that collisions within a cluster become near
elastic. Elastic collisions tend to smoothen out density inhomogeneities.
Therefore, clusters become less compact and
fragmentation is initiated. The observations are
mathematically summarised as 
\bea \sigma(t,\delta) &\simeq& \delta^2
f_1\left(t \delta^{2/x_1}\right), ~\mathrm{fixed}~
m, \label{f1scaling}\\ \sigma(t,m) &\simeq& f_2\left(\frac{t}{
  m^{x_2/x_1}}\right), ~\mathrm{fixed}~\delta,
\label{f2scaling}
\eea
where $x_1$, $x_2$ are scaling exponents and $f_1$, $f_2$ scaling functions
such that  $f_1(z) \sim z^{-x_1}$, $z \ll 1$, and $f_2(z) 
\sim z^{-x_1}$, $z \ll 1$. 
Thus, for fixed $\delta$, $\sigma \sim m^{x_2} t^{-x_1}$
for initial times.

\begin{figure}
\includegraphics[width=\columnwidth]{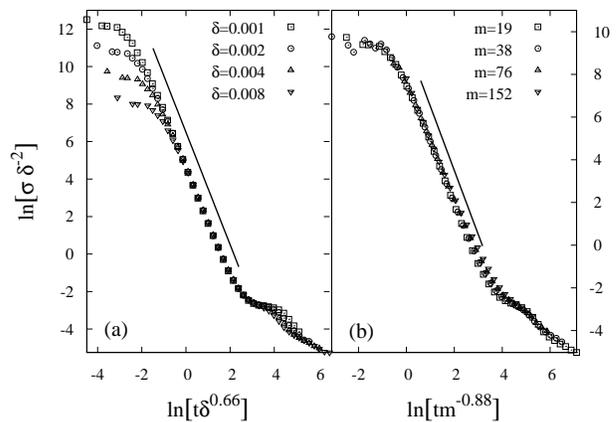}
\caption{\label{sigma_scaling} Data collapse when scaled velocity
fluctuations $\sigma \delta^{-2}$ is plotted against (a) scaled time variable 
$t \delta^{0.66}$ and (b) scaled time variable $t m^{-0.88}$. 
The data are for (a) 
$m=19$ and (b) $\delta=0.004$. The solid straight lines have slope
$=-3.0$.
}
\end{figure}

The exponents $x_1$, $x_2$ may be obtained from the data collapse of the data
in Fig.~\ref{sigma_raw} when scaled as in Eqs.~(\ref{f1scaling}) and
(\ref{f2scaling}). The scaled data is shown in Fig.~\ref{sigma_scaling}(a)
[Eq.~(\ref{f1scaling})] and Fig.~\ref{sigma_scaling}(b) 
[Eq.~(\ref{f2scaling})]. From these, we obtain
\begin{subequations}
\label{eq:exponents}
\bea
x_1 &=& 3.00\pm 0.06,\\
x_2 &=& 2.66\pm 0.08
\eea
\end{subequations}
Note that these values of $x_1$ and $x_2$
imply that the crossover time $t_1$, relevant for $\sigma$,
scales as $\delta^{-2/x_1} \sim \delta^{-0.66}$. This time scale is much
smaller that $t_2\sim\delta^{-3}$, which is the crossover time associated
with velocities of nearly all particles becoming of order $\delta$,
i.e. all collisions becoming near elastic.

\begin{figure}
\includegraphics[width=\columnwidth]{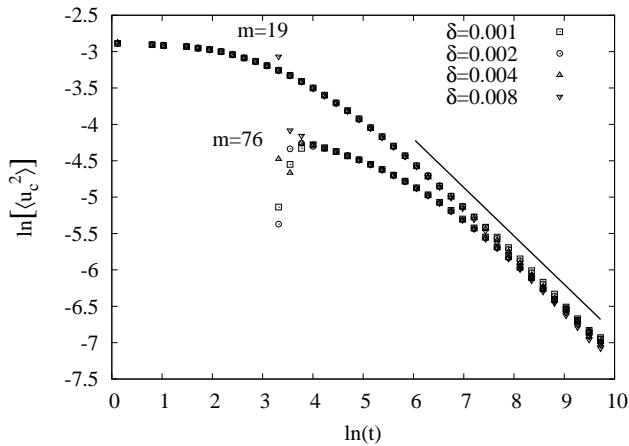}
\caption{\label{energy_raw} The square of the centre of mass velocity 
$\langle u_c^2 \rangle$ of a cluster of size $m$ as a function of time $t$ 
for different $\delta$. The data are for sizes $m=19$ and $m=76$. For the times
shown, there is no dependence on $\delta$. The solid line has slope $-2/3$.
}
\end{figure}

To contrast the above intra-cluster velocity fluctuations $\sigma$
with typical cluster velocities, 
we study the centre of mass velocity ${\bf u}_c$ of a cluster.  It
is well known that average energy per particle in the cooling gas
decreases as $t^{-2/3}$ \cite{bennaim1}. Not surprisingly, we find
that the $\langle u_c^2 \rangle$ for a cluster decays with the same law.  
In Fig.~\ref{energy_raw}, we show the variation of $\langle u_c^2 \rangle$
with time for different $\delta$ and two different cluster sizes.
$\langle u_c^2 \rangle$ decreases as $t^{-{2/3}}$ at large $t$.
There is no signature of any crossover
across any intermediate time scale $t_1$ nor any dependence on
$\delta$. The intra-cluster near elastic 
collisions affect velocity fluctuations but not the typical speeds, which are
affected only by cluster--cluster collisions.  

\begin{figure}
\includegraphics[width=\columnwidth]{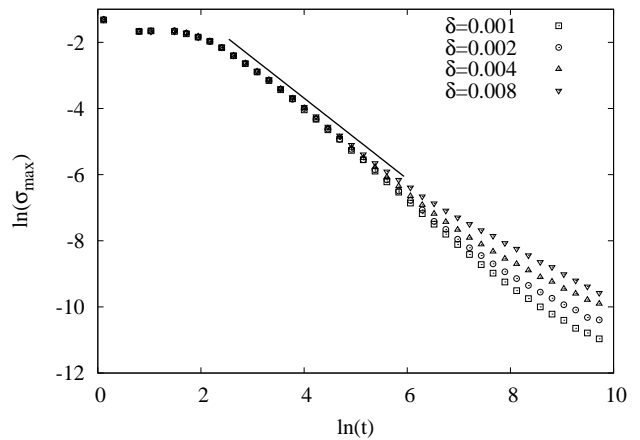}
\caption{\label{largest_mass} The velocity fluctuations $\sigma_{max}$
of the largest cluster, as a function of time $t$ for different
$\delta$. For short times the curves are independent of $\delta$. 
Deviation from power law is seen earlier for larger $\delta$. The
solid straight line has slope $-1.22$.
}
\end{figure}

We provide a further check for the exponent values in Eq.~(\ref{eq:exponents})
by quantifying the velocity fluctuations $\sigma_{\mathrm{max}}$ 
of the largest cluster in the system. From Eq.~(\ref{f2scaling}), we obtain
that $\sigma_{\mathrm{max}} \sim M_{\mathrm{max}}^{x_2} t^{-x_1}$, where
$M_{\mathrm{max}}(t)$ is the size of the largest cluster at time $t$. Noting
that $M_{\mathrm{max}}(t) \sim t^{2/3}$ \cite{frachebourg1999,bennaim1}, 
we obtain $\sigma_{\mathrm{max}} \sim
t^{2 x_2/3 -x_1} \sim t^{-1.22}$, where we substituted the values of $x_2$ and
$x_1$ from Eqs.~(\ref{f1scaling}) and (\ref{f2scaling}). In
Fig.~\ref{largest_mass}, we show the variation of $\sigma_{\mathrm{max}}$ with
time $t$ for different values of $\delta$. The temporal regime which is
independent of $\delta$ is consistent with the exponent $1.22$.

We now show that the above crossover of $\sigma$ from a power law is
linked very closely to the initiation of fragmentation in clusters.
The fragmentation rate $W_f(m,t)$ defined in Eq. (\ref{eq:frag_defn})
is numerically measured as follows.  At time $t$, all the clusters of
a particular size $m$ are identified. At time $t+\Delta t$, the
fraction of the identified clusters whose size has reduced is
calculated. That fraction is equal to $W_f(m,t) \Delta t$.
In the simulations, we choose $\Delta t$ to be one, so that sufficient
statistics may be obtained.

\begin{figure}
\includegraphics[width=\columnwidth]{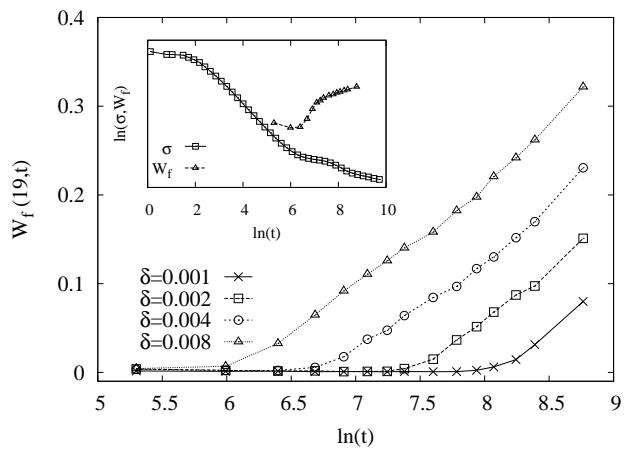}
\caption{\label{frag_time_sigma} Fragmentation rate $W_f$ as a function of
time $t$ for different $\delta$. The data are for cluster size $19$.
Inset: Velocity fluctuations $\sigma$ and $W_f$ as a function of time.
Increase in fragmentation rate coincides with saturation of velocity
fluctuations. The data are for $\delta=0.004$ and cluster size $19$.
}
\end{figure}

In Fig.~\ref{frag_time_sigma}, we show the fragmentation rate $W_f$
for cluster size $19$ for
different values of $\delta$. There is a sharp increase in the
fragmentation rate, with the increase setting in earlier for larger
$\delta$.  In the inset of Fig.~\ref{frag_time_sigma}, we superimpose
the fragmentation rate on the plot of $\sigma$ with time for the same
value of $\delta$ and cluster size. Clearly, the increase in
fragmentation rate coincides with the deviation from power law
behaviour of the velocity fluctuations.  This increased fragmentation
rate results in fluctuation dominated coarsening and ultimately to
the breakdown of Porod law \cite{shinde1}.

\begin{figure}
\includegraphics[width=\columnwidth]{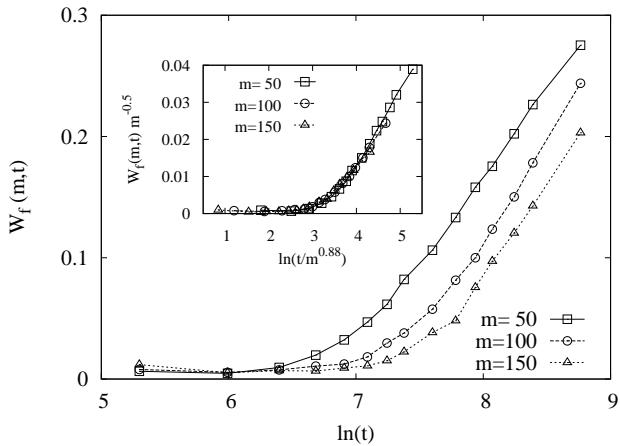}
\caption{\label{frag_rate_mass} Fragmentation rate $W_f$ as a function of
time $t$ for different cluster sizes $m$. The data are for $\delta=0.008$.
Inset: Data collapse when  $W_f$ and time are scaled as in
Eq.~(\ref{fragscaling}) with $\eta=0.5$.
}
\end{figure}

The fragmentation rate $W_f$ depends on the cluster size too. In
Fig.~\ref{frag_rate_mass}, we show the variation of $W_f$ with $t$ for
different $m$ and fixed $\delta$. 
The crossover time beyond which increased fragmentation
is seen increases with $m$. Knowing that $t \sim m^{x_2/x_1}$
(Eq. (\ref{f2scaling})), we write \be W_f (t,m) \simeq m^{\eta}
f_w\left(\frac{t}{ m^{x_2/x_1}}\right), ~\mathrm{fixed}~\delta,
\label{fragscaling}
\ee
where $\eta$ is an unknown exponent. We see that for $\eta=0.5$, we obtain
excellent data collapse [see inset of Fig.~\ref{frag_rate_mass}].

\begin{figure}
\includegraphics[width=\columnwidth]{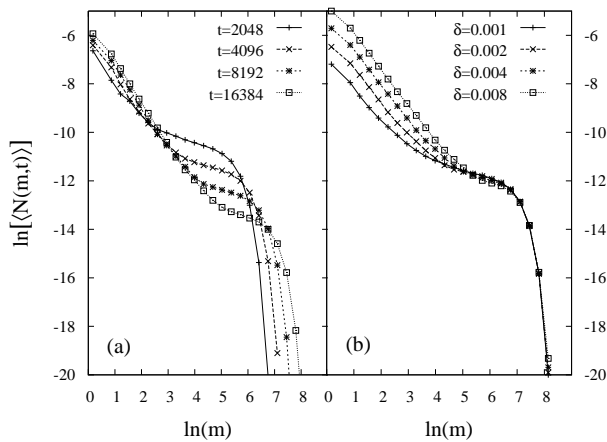}
\caption{\label{prob_raw} The average cluster distribution $\langle
N(m,t) \rangle $ as a function of cluster size $m$ for (a) fixed
$\delta=0.004$, different times and (b) different $\delta$, fixed time
$t=16384$.
}
\end{figure}

An increased fragmentation rate will result in clusters up to a size
$m^*(t,\delta)$ breaking apart. We believe that this is the origin of
the breakdown of Porod law, as fragmentation results in new structures
at small scales. One way to capture this is to study the average cluster size
distribution $\langle N(m,t) \rangle$, where $N(m,t)$ is the number of clusters
of size $m$ at time $t$ and the average is over space
and histories. We would like to investigate whether, even in the presence of
fragmentation, the cluster size distribution can be described by the sticky
gas.  We argue below that while some regimes of $\langle N(m,t) \rangle$
resemble the sticky gas, other regimes differ, but their scaling may be 
obtained from that of $\sigma(m,t)$.
The mean cluster size distributions
are shown for different times in Fig.~\ref{prob_raw}(a) and for
different $\delta$ in Fig.~\ref{prob_raw}(b). 

\begin{figure}
\includegraphics[width=\columnwidth]{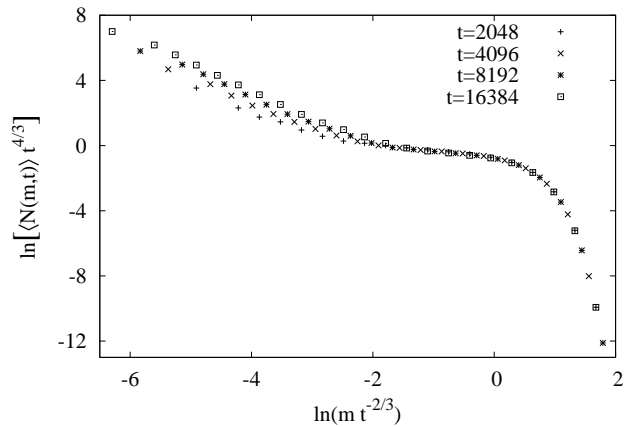}
\caption{\label{prob_large_scaling} Data collapse for large cluster
sizes  $m$
when the cluster distribution $\langle N(m,t) \rangle$ and $m$ are
scaled as in Eq.~(\ref{stickyscaling}). The data are for
$\delta=0.004$.
}
\end{figure}

We note that for fixed time,
$\langle N(m,t) \rangle$ for large masses has no dependence on $\delta$ [see
Fig.~\ref{prob_raw}(b)]. 
Thus, we expect that for masses greater than a mass
cutoff $m^*(t,\delta)$, fragmentation is not relevant and $\langle N(m,t)
\rangle$ should have the same scaling behaviour as in the sticky gas.
For the sticky gas, 
it is known \cite{frachebourg1999} that
\be 
\langle N(m,t) \rangle \simeq \frac{1}{t^{4/3}} f_3
\left( \frac{m}{t^{2/3}} \right), 
\label{stickyscaling}
\ee 
where the scaling function $f_3(z) \sim z^{-1/2}$, $z \ll 1$ and
$f_3(z) \rightarrow 0$ for $z \gg 1$. For masses $m > m^*(t,\delta)$ for
which fragmentation is not important, we confirm numerically that the same
scaling holds.
In Fig.~\ref{prob_large_scaling}, we scale the data of
Fig.~\ref{prob_raw}(a) as in Eq.~(\ref{stickyscaling}) and we see
excellent data collapse for large cluster sizes, confirming that
fragmentation can be neglected for cluster sizes larger than $m^*(t,\delta)$. 

\begin{figure}
\includegraphics[width=\columnwidth]{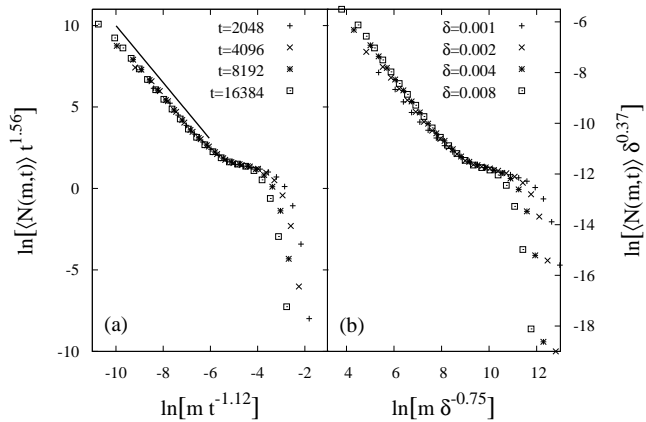}
\caption
{\label{prob_small_scaling} Data collapse for small cluster 
sizes $m$ when the cluster distribution $\langle N(m,t) \rangle$ and $m$ 
are scaled as in (a) Eq.~(\ref{smallmasstime}) for fixed $\delta=0.004$ 
and (b) Eq.~(\ref{smallmassdelta}) for fixed $t=16384$. The solid line 
has slope $1.75$.
}
\end{figure}

We note that in Fig.~\ref{prob_large_scaling}, there is no data collapse 
for small cluster sizes, when the data is scaled as in 
Eq.~(\ref{stickyscaling}), and thus small cluster sizes 
cannot be described by the sticky 
gas scaling. We argue that its scaling can be obtained from that of the scaling 
of $\sigma$. For a fixed $\delta$, varying $t$, $\langle N(m,t) \rangle $ 
should have the scaling form
\be
\langle N(m,t) \rangle \simeq \frac{1}{t^{\alpha}} f_4 \left(
\frac{m}{t^{x_1/x_2}}
\right), \quad m \ll m^*, ~ \mathrm{fixed}~\delta,
\label{smallmasstime}
\ee 
where $\alpha$ is an exponent which we determine by examining the
large $z$ behaviour of the scaling function $f_4(z)$. For large $z$,
$f_4(z)$ should be such that it crosses over to the small $z$ behaviour
of $f_3(z)$. Thus $f_4(z) \sim z^{-1/2}$ for $z \gg 1$. Comparing the
time dependence of Eqs. (\ref{stickyscaling}) and
(\ref{smallmasstime}) in the latter limit, we obtain \be \alpha = 1+
\frac{x_1}{2 x_2} \approx 1.56.
\label{eq:alpha}
\ee
The data when scaled as in Eqs.~(\ref{smallmasstime}) and (\ref{eq:alpha}) 
with $x_1$ and $x_2$ as in Eq.~(\ref{eq:exponents}) is shown in 
Fig.~\ref{prob_small_scaling}(a). We obtain good data collapse for the
small cluster sizes, 
showing that knowing the scaling behaviour of $\sigma$ helps us
obtain the scaling behaviour of $\langle N(m,t) \rangle$. The small $z$
behaviour of $f_4(z)$ can be determined numerically. We find that $f_4(z)
\sim z^{-\tau}$ with $\tau = 1.75 \pm 0.08$ [see solid line in
Fig.~\ref{prob_large_scaling}(a)].

The scaling of the small cluster sizes with $\delta$ can also be obtained
from the scaling of $\sigma$. Knowing that $m^*$ scales as $\delta^{2/x_2}$,
we write
\be
\langle N(m,\delta) \rangle \simeq \frac{1}{\delta^{\beta}} f_5 \left(
\frac{m}{\delta^{2/x_2}} \right), \quad m \ll m^*,~ \mathrm{fixed}~t,
\label{smallmassdelta}
\ee
where the exponent$\beta$ can be determined as above by constraining the large
$z$ behaviour of the scaling function $f_5(z)$ to be the same as the small $z$
behaviour of $f_3(z)$. This immediately implies that $f_5(z) \sim z^{-1/2}$
for $z \gg 1$ and
\be
\beta = \frac{1}{x_2} \approx 0.37
\label{eq:beta}
\ee
The data for cluster size distribution when scaled as in 
Eqs.~(\ref{smallmassdelta}) and (\ref{eq:beta})
with $x_1$ and $x_2$ as in Eq.~(\ref{eq:exponents}) is shown in
Fig.~\ref{prob_small_scaling}(b). We obtain reasonable data collapse for the
small cluster sizes. However, given the range and quality of data, it is
possible to obtain data collapse for a range of $x_2$.

\section{\label{sec:summary}Discussion}

To summarise, we studied velocity fluctuations and size 
distribution of clusters in a freely cooling granular gas in one 
dimensional ring evolving via ballistic motion and inelastic collisions. 
The coefficient of restitution was $r_0<1$ for relative velocity greater 
than $\delta$ and $1$ otherwise. The aim of the paper was to understand 
the consequences of a non-zero $\delta$ on the structure of clusters for 
large times.

For granular gases with realistic velocity dependent coefficient 
of restitution, it was recently shown \cite{shinde1, shinde2}  
that the nature of coarsening in the  inhomogeneous cooling regime 
is not the same at all 
times. Beyond some crossover scale $t_1$, the coarsening behaviour 
changes at the macroscopic level from one that obeys Porod law to one that
violates Porod law. These interesting numerical findings lacked
a mesoscopic explanation of how and why the crossover occurs. The current
paper provides an explanation. We 
demonstrate in this paper, that the transition from sticky gas regime to 
fluctuation dominated ordering regime within the inhomogeneous cooling regime
may be viewed as a 
growing dominance of an underlying fragmentation process competing 
against the dominant clustering process. The fact that clusters break 
up is shown by the change of behaviour of the 
variance of particle velocities within a cluster. This crossover in velocity
fluctuations coincide with an increase in the fragmentation rate of clusters
leading to a richer fine structure reflected in the density--density
correlations. 

The velocity fluctuations within a cluster were found to decrease as a 
power law with time, with the velocity fluctuations being much smaller than
the centre of mass velocities. However, when these fluctuations became of order 
$\delta$, then intra cluster collisions became mostly elastic and the 
clusters start to fragment. This emergent phenomena was quantified by 
defining an effective fragmentation rate for a cluster. The 
fragmentation rate was seen to rise sharply at some cluster size 
dependent time with the crossover time increasing with decreasing 
$\delta$. Once fragmentation sets in, the cluster size distribution 
$\langle N(m,t) \rangle$ changes drastically from that of the sticky gas 
($r_0=0$, $\delta=0$). However, the scaling of $\langle N(m,t) \rangle$ 
could be related to that of the velocity fluctuations $\sigma(m,t)$ 
which was completely characterised by two independent exponents $x_1$ and
$x_2$. It was 
also observed that the total energy of the system as well as clusters 
continue to decay as $t^{-2/3}$, showing no signature of the structural 
changes in the clusters.

We believe that many of these results (qualitative) will be carried over 
to higher dimensions. In two dimensions, for coarse grained velocities, 
it was shown \cite{bennaim2} that the velocity fluctuations scale 
differently from the typical velocity. Hence, we expect a crossover when
these fluctuations become comparable to $\delta$, and thus fragmentation to 
be relevant for two dimensions too, and consequently a regime where 
coarsening is fluctuation dominated. It would be interesting to verify 
it numerically.

The velocity scale $\delta$ is relevant and not just a computational 
tool. Experimentally,  $r(v)$ approaches $1$ when the relative velocity $v$
tends to zero, i.e., $r(v) \simeq 1- (v/\delta)^\chi+ \ldots$, $v/\delta \ll
1$. The exponent $\chi$ takes a variety of values. Within the framework of
viscoelastic theory, $\chi=1/5$. Systems with $\chi<1$ cannot be studied 
using event driven molecular dynamics simulations as inelastic collapse
prevents the simulation from proceeding forward. It would be interesting to
use conventional molecular dynamics simulations to verify whether the
observations of this paper is valid for $\chi<1$.

Another question of interest is the construction of lattice 
models that reproduce the coarse grained behaviour of the granular gas. 
Such models are not only computationally much faster, but also may be 
the first steps towards building effective field theories for the 
system. In a recent paper \cite{dey2011}, a stochastic lattice model was 
studied which reproduced all features of the sticky gas. It would be 
interesting to see whether fragmentation can be incorporated into this 
lattice model such that the coarse grained behaviour seen in this paper 
is reproduced.

\begin{acknowledgments}
MS was partially supported by the International Young Scientist Fellowship of
Institute of Physics, Chinese Academy of Sciences under the Grant No.~2011001.
A part of this work (MS) was carried out at Centre for Nonlinear Studies, 
Hong Kong Baptist University.
\end{acknowledgments}

\end{document}